\documentclass{iopart}
\usepackage{graphicx}
\usepackage{color}

\newcommand{\PRB}{\textit{Phys.\ Rev.\ B }}

\begin{document}

\title{Molecular neuron based on the Franck-Condon blockade}

\author{C Timm$^1$ and M Di Ventra$^2$}

\address{$^1$ Institute of Theoretical Physics, Technische Universit\"at
Dresden, 01062 Dresden, Germany\\$^2$ Department
of Physics, University of California, San Diego, La Jolla, CA 92093,
USA}
\eads{\mailto{carsten.timm@tu-dresden.de}, \mailto{diventra@physics.ucsd.edu}}

\begin{abstract}
Electronic realizations of neurons are of great interest as building blocks for
neuromorphic computation. Electronic neurons should send signals into the input
and output lines when subject to an input signal exceeding a given threshold, in
such a way that they may affect all other parts of a neural network. Here, we
propose a design for a neuron that is based on molecular-electronics components
and thus promises a very high level of integration. We employ the Monte Carlo
technique to simulate typical time evolutions of this system and thereby show
that it indeed functions as a neuron.
\end{abstract}




\section{Introduction}

Neurons are the fundamental building blocks of information processing by the
nervous system of every multicellular animal on Earth. They are cells of various
degrees of sophistication serving the ultimate goal of processing and
transferring information to adjacent neurons, and thus coordinating all of the
animal's activities.

Reproducing some, if not all, of the capabilities of a biological neural
network using electronic components has always been a goal of research into
artificial intelligence~\cite{book_AI}. The benefits would range from
massively parallel unconventional computing~\cite{maze} to electronics that
adapts to the type of signal that is fed into the
network~\cite{adaptiveo,assmem}, to name just a few.
Traditional solid-state transistors can indeed mimic the main characteristics
of a neuron (see, e.g.,~\cite{masson}). However, they suffer from integration
limits and therefore are unlikely to reach the scalability of biological brains.
In this respect, neurons based on molecular electronics appear to be a viable
alternative. For instance, recent work has demonstrated an organic-nanoparticle
transistor that operates as a spiking neuron~\cite{alibart}.
However, the design~\cite{alibart} relies on a combination of a thin film of
pentacene molecules and gold nanoparticles. To simplify its fabrication and
increase its integration capabilities it would be desirable to instead design
an all-molecular circuit that operates as a neuron. Such a neuron could serve
as a building block for highly-integrated neural computers~\cite{memco}.

In the present work we propose a type of molecular neuron based on the
mechanism of Franck-Condon blockade
\cite{KoO05,KRO05,KOA06,DGR06,LeW08,HuB09,LSI09,CML10,DoT12} and show that it
reproduces all the main features of a spiking neuron. Our
guiding principles are the following: {(i)} the design should be as simple as
possible, and {(ii)} it should not require excessive fine tuning of
parameters in order to function.
To start with, we have to specify how we want the device to behave. Typical
molecular-electronics components suggested in the literature pump electrons
between source and drain electrodes, controlled by a gate voltage. Hence, the
input signal for the artificial neuron will likely be a current. Furthermore, we
require the neuron to fire sharp voltage spikes into its output line but also
into its input line when the input current exceeds a certain threshold, so that
it may affect all other parts of the network~\cite{assmem}. Below the threshold,
the neuron should be quiescent.

We thus need a molecular device that generates sharp spikes. Our central idea
is to use a molecular transistor in which the electrons are strongly coupled to
a vibrational mode. This coupling can lead to \emph{Franck-Condon} (FC)
\emph{blockade}
\cite{KoO05,KRO05,KOA06,DGR06,LeW08,HuB09,LSI09,CML10,DoT12}. The essential
physics of this phenomenon is the following: The relevant
vibrational mode is described by a normal coordinate $x$. The potential
(deformation) energy $V_\mathrm{def}(x)$
is a different function of $x$ for different charge states.
In particular, for strong electron-vibron coupling, the value $x=x_\mathrm{min}$
minimizing $V_\mathrm{def}(x)$ shows a large shift between charge states. The
molecule is in some vibrational state at time $t$, described
by a wave function $\psi(x,t)$. Let us assume that it is in the vibrational
ground state. Now, when an electron tunnels in or out, the potential energy
$V_\mathrm{def}(x)$ is suddenly switched to a quite different function, with
shifted minimum. Hence, the molecule finds itself in a state
far from the vibrational ground state for the new potential. More
precisely, the overlap between the old and new ground states is very small. The
overlap between other low-energy eigenstates with respect to the two potentials
is also suppressed. These overlaps are called FC matrix elements
\cite{MAM04,KoO05,KOA06,LeW08,DoT12}. Due to the small FC matrix elements,
electronic tunneling transitions that are energetically possible can
nevertheless be strongly suppressed, in particular for low bias voltages.
Consequently, the current through a molecular device with strong electron-vibron
coupling is suppressed at low bias. This effect is called Franck-Condon
blockade.

The dynamics in the FC regime is also unusual:
The electrons tunnel in avalanches separated by quiescent intervals
\cite{KoO05,KRO05}. The reason is that while the probability for an electron
to tunnel into the molecular orbital is very small, when it finally does so and
then
tunnels out again, the molecule likely returns to a higher vibrational level
since the FC matrix element for such a process is larger than for a return into
the ground state. But this makes it much more likely for another
electron to tunnel. Thus we get an avalance, until the system finally drops
back into the vibrational ground state. Our idea is to use these avalanches to
generate spikes when the neuron is active.

The remainder of this paper is organized as follows: In section
\ref{sec.circuit}, we introduce the molecular circuit and
qualitatively discuss its behavior. The model and simulation technique are
discussed in section \ref{sec.simulation}. In section \ref{sec.results} we
present results for the time-dependent behavior of the neuron. We summarize the
paper in section \ref{sec.summary}.

\section{Molecular circuit}
\label{sec.circuit}

\begin{figure}
\centerline{\includegraphics[scale=0.55,clip]{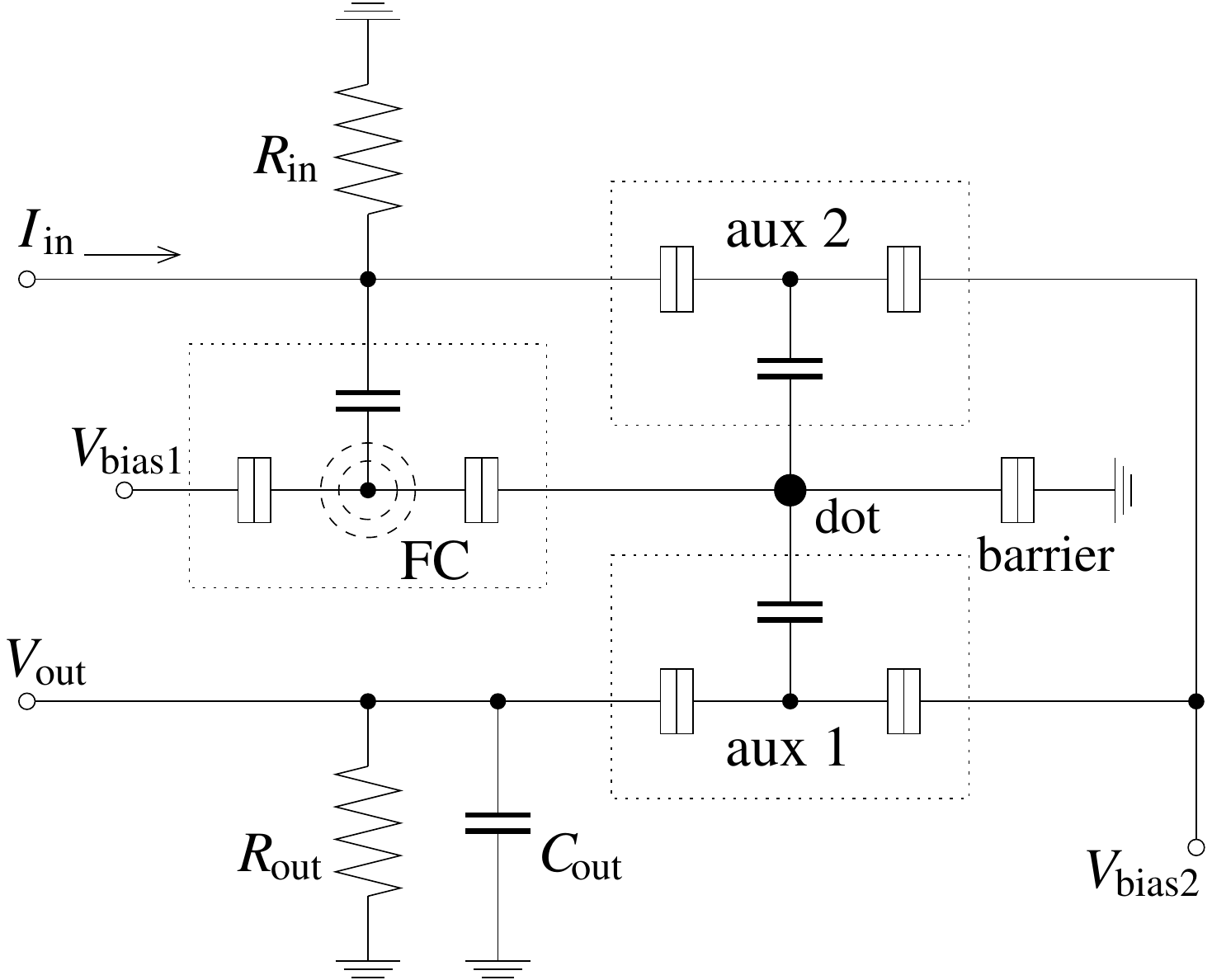}}
\caption{Molecular neuron circuit
based on FC blockade. The three dotted rectangles
enclose single-molecule devices, as explained in the text. The black circle
marked ``dot'' denotes a quantum dot separated from ground by a tunneling
barrier.\label{fig.circuit}}
\end{figure}

The circuit diagram is shown in figure \ref{fig.circuit}.
The active part of the circuit is a single-molecule transistor in the FC
regime (dotted rectangle labeled ``FC'' in figure \ref{fig.circuit}).
Disregarding the transistor labeled ``aux 2'' for the time being, the input
current $I_\mathrm{in}$
leads to a voltage drop across the resistor $R_\mathrm{in}$. This voltage is
applied to the gate of the FC molecular transistor. Thereby, the input current
is able to switch this transistor between OFF and ON states. The OFF state
does not have any molecular transition energies between the electrochemical
potentials of the source and drain electrodes so that sequential tunneling is
thermally suppressed. On the other hand, in the ON state such transitions
exist, but are suppressed by small FC matrix elements. In the ON state, there
is thus a current from the source electrode kept at fixed voltage
$V_\mathrm{bias\:1}$ to the drain electrode. As noted above, the current flows
in avalanche-like bursts \cite{KoO05,KRO05}.

Any avalanche dumps a relatively large number of carriers within a short time
into a quantum dot attached to the drain (circle labeled ``dot'' in figure
\ref{fig.circuit}). This quantum dot may be realized by a small metallic cluster
or a large organic molecule. The carriers leave the dot via another tunneling
barrier (labeled ``barrier'' in figure \ref{fig.circuit}).
The charge on the dot is coupled to the gate of two auxiliary molecular
transistors, labeled ``aux 1'' and ``aux 2'' in figure \ref{fig.circuit}. They
are not in the FC regime and will in the following be modeled by a single
orbital with strong Coulomb repulsion but not coupled to vibrations. The first
auxiliary device controls current flowing from a source electrode at fixed
voltage $V_\mathrm{bias\:2}$ to a drain electrode connected to the output line.
When the dot is charged due to the FC molecule being active, a current can flow.
By choosing large tunneling rates to the source and drain contacts, the device
can, in principle, act as an amplifier. The current flows through a resistor
$R_\mathrm{out}$ to the ground, creating a voltage drop, which we use as
the output
signal (in the simulation, we will assume that the output line is current free).
During an avalanche, the auxiliary device is active, a
current flows, and there is a nonzero voltage drop. The result is a voltage
appearing only in short time intervals, as required. We introduce a capacitor
$C_\mathrm{out}$ to ground to dampen the sharp voltage spikes, which would
otherwise be $\delta$-function-like. In a real setup, intrinsic capacitances
would always lead to such a broadening.

The second auxiliary device (``aux 2'' in figure \ref{fig.circuit}) essentially
works like the first one, except that it inserts a current into the input line,
when the FC molecule is active. The current leads to a voltage drop across the
input resistor $R_\mathrm{in}$. The voltage spikes are dampened by the gate
capacitor of the FC molecule. We find that an additional capacitor is not
required. The
connection of both the gate of the FC molecule and the drain of the second
auxiliary molecule to the same input line of course leads to feedback. If the
feedback is too strong, a current avalanche through the FC molecule
could be choked off immediately. However, we will show below that this
undesirable behavior can be avoided by a suitable choice of circuit parameters.

\section{Simulation method}
\label{sec.simulation}

In the field of transport through molecular devices, the choice of the
theoretical approach depends on details of the system: If the
hybridization between molecular orbitals and states in the electrodes is
relatively weak so that it can be treated perturbatively, master-equation
approaches \cite{ScS94,BrF03,MAM04,KoO05,KOA06,DGR06,LeW08,Tim08,HuB09}
are appropriate. The strengths of interactions, such as the Coulomb repulsion
between electrons in the molecule and the electron-vibron coupling, may be large
in these approaches. On the other hand, if interactions are weak, methods based
on non-equilibrium Green functions can be employed~\cite{book,STB05}. These
approaches are able to treat the hybridizations exactly and are thus not
limited to small hybridizations. For our purposes, however, Green-function
methods are not suitable since our neuron circuit relies on the strong
electron-vibron interaction in the FC molecule.

\subsection{Model}

In this work, we assume the hybridizations to be weak and treat them in
leading-order perturbation theory, i.e., in the sequential-tunneling
approximation. We also assume that dephasing is
rapid so that off-diagonal components in the reduced density matrices
(coherences) of the molecular transistors can be ignored~\cite{book}. The
sequential-tunneling rates are then well known
\cite{MAM04,KoO05,KOA06,LeW08,DoT12}.

\begin{figure}
\begin{center}
(a)\includegraphics[scale=0.55,clip]{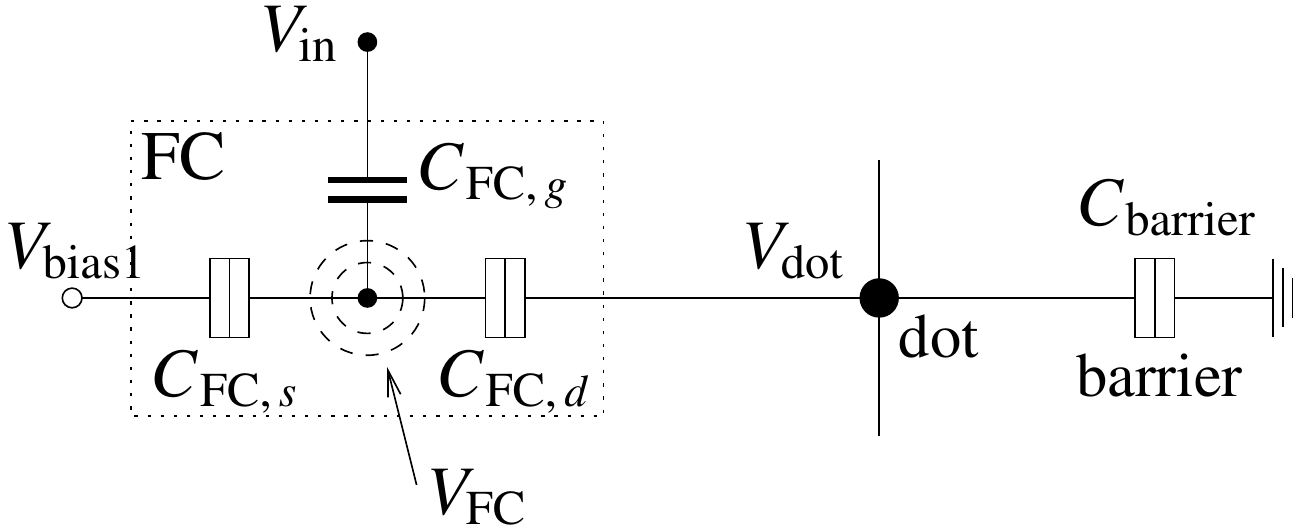}\\
(b)\includegraphics[scale=0.55,clip]{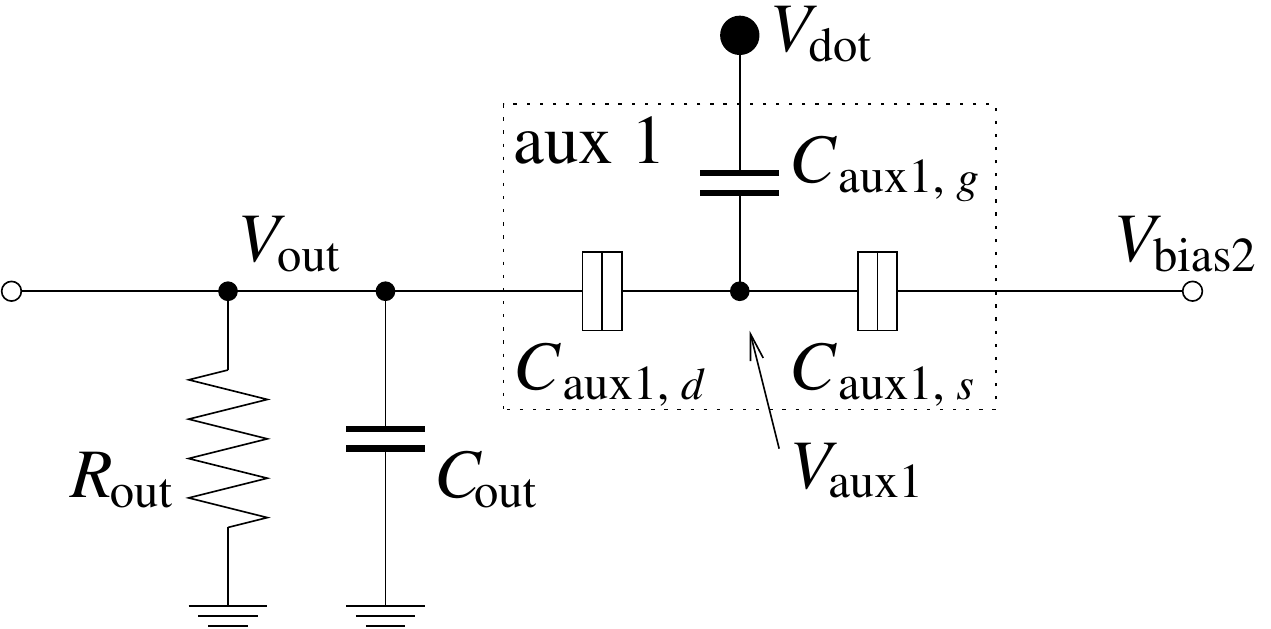}\\
(c)\includegraphics[scale=0.55,clip]{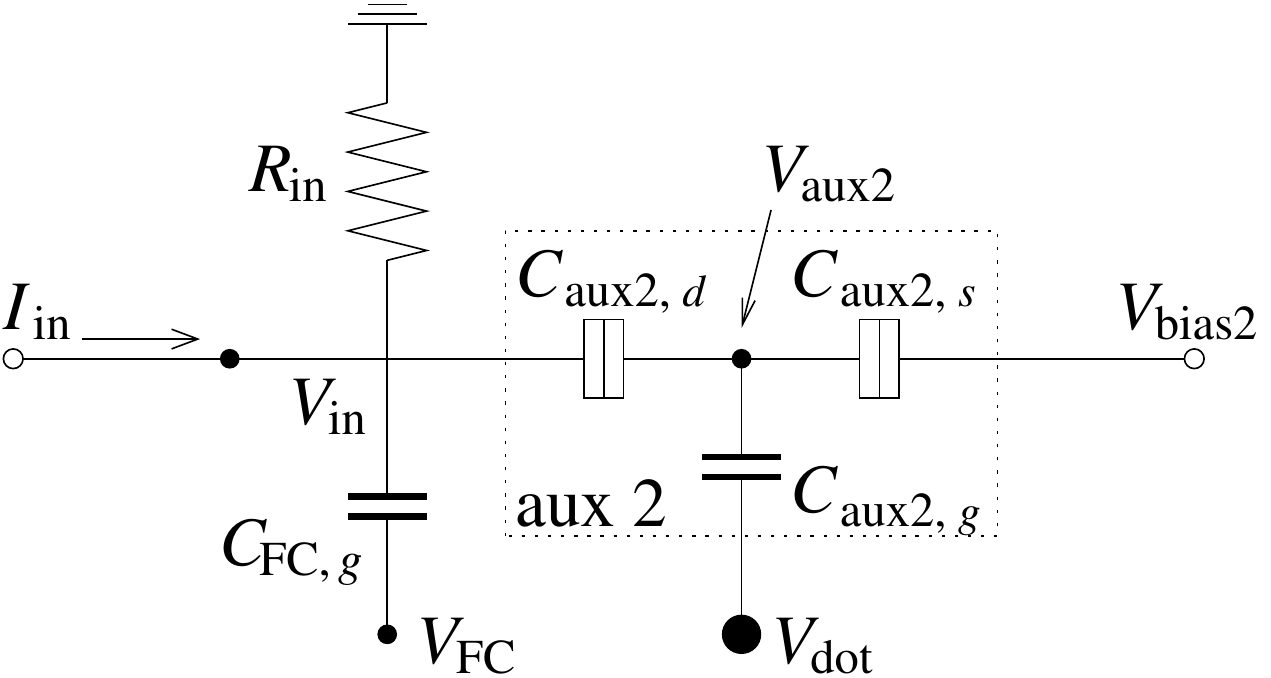}
\end{center}
\caption{Details of the circuit diagram of the molecular neuron, indicating
quantities used in the text. (a) FC molecule, dot, and tunneling barrier. (b)
Auxiliary molecule 1. (c) Auxiliary molecule 2.\label{fig.details}}
\end{figure}

We take all molecular transistors to contain a single relevant molecular
orbital. The Hamiltonian of the FC molecule reads
\begin{eqnarray}
H_\mathrm{FC} & = & \sum_\sigma (\epsilon_\mathrm{FC} - eV_\mathrm{FC})\,
  d^\dagger_\sigma d_\sigma
  + U_\mathrm{FC}\, d^\dagger_\uparrow d_\uparrow d^\dagger_\downarrow
  d_\downarrow \nonumber \\
&& {} + \hbar\omega_v\, \left(b^\dagger b + \frac{1}{2}\right)
  + \lambda \hbar\omega_v\, (b+b^\dagger) \sum_\sigma d^\dagger_\sigma d_\sigma
  ,
\label{HFC.1}
\end{eqnarray}
where $d^\dagger_\sigma$ creates an electron of spin $\sigma$ in the molecular
orbital and $b^\dagger$ is the creation operator of a harmonic vibrational mode.
The energy $\epsilon_\mathrm{FC}$ of the molecular orbital is shifted by the
electric potential $V_\mathrm{FC}$. The potential is coupled to
the voltages $V_\mathrm{bias\:1}$, $V_\mathrm{dot}$, and $V_\mathrm{in}$
applied to the source, drain, and gate electrodes, respectively, through the
capacitances of these contacts, as further discussed below.
The voltages and capacitances are indicated in figure \ref{fig.details}(a).
$\hbar\omega_v$ in (\ref{HFC.1}) is the energy quantum of the harmonic
oscillator and the dimensionless constant $\lambda$ describes the strength of
the electron-vibron coupling.

The eigenstates of the isolated molecule are
written as $|nq\rangle$, where the electronic state is denoted by
$n=0,{\uparrow},{\downarrow},{\uparrow\downarrow}$ for the empty, spin-up,
spin-down, and doubly occupied state, respectively. The vibrational states are
enumerated by the harmonic-oscillator quantum number $q=0,1,2,\ldots$
The eigenenergies read \cite{DoT12}
\begin{eqnarray}
E^\mathrm{FC}_{nq} & = & (\epsilon_\mathrm{FC} - eV_\mathrm{FC} - \lambda^2
\hbar\omega_v)\,  n_d \nonumber \\
&& {}+ \left( \frac{U_\mathrm{FC}}{2} - \lambda^2 \hbar\omega_v\right)\,
  n_d(n_d-1)
  + \hbar\omega_v\, \left(q + \frac{1}{2}\right) ,
\end{eqnarray}
where $n_d=0,1,2$ is the number of electrons in the state $|nq\rangle$.

The sequential-tunneling rates from state $|n'q'\rangle$ to state $|nq\rangle$
involving an electron tunneling out of the molecule into the source or the drain
electrode are
\begin{equation}
R^{\mathrm{out},s}_{n'q' \to nq} = \Gamma_\mathrm{FC}\,
  f(E_{nq} - E_{n'q'} - eV_\mathrm{bias\:1})
  \sum_\sigma |D^\sigma_{nn'}|^2\, |F_{qq'}|^2
\label{RFC.1a}
\end{equation}
and
\begin{equation}
R^{\mathrm{out},d}_{n'q' \to nq} = \Gamma_\mathrm{FC}\,
  f(E_{nq} - E_{n'q'} - eV_\mathrm{dot})
  \sum_\sigma |D^\sigma_{nn'}|^2\, |F_{qq'}|^2 ,
\label{RFC.1b}
\end{equation}
respectively, where $\Gamma_\mathrm{FC}$ is a bare tunneling rate determined by
the hybridization and the density of states in the electrodes,
$f(x)$ is the Fermi function, $D^\sigma_{nn'} \equiv \langle
n| d_\sigma |n'\rangle$ are matrix elements of the electronic annihilation
operator, and
\begin{equation}
|F_{qq'}|^2 = \frac{q_<!}{q_>!}\:
  \lambda^{2(q_>-q_<)}\, e^{-\lambda^2}\,
  \big[ L^{q_>-q_<}_{q_<}(\lambda^2) \big]^2
\end{equation}
are FC matrix elements squared \cite{MAM04,KoO05,KOA06,LeW08,DoT12}. Here, $q_<
\equiv \min(q,q')$, $q_> \equiv \max(q,q')$, and $L^i_j(x)$ are generalized
Laguerre polynomials. In the interest of a simple model, we assume the bare
tunneling rate $\Gamma_\mathrm{FC}$ to be the same for source and drain. This
assumption is not crucial for a functioning neuron. The corresponding rates for
an electron tunneling \emph{into} the molecule read
\begin{equation}
R^{\mathrm{in},s}_{n'q' \to nq} = \Gamma_\mathrm{FC}\,
  f(E_{nq} - E_{n'q'} + eV_\mathrm{bias\:1})
  \sum_\sigma |D^\sigma_{n'n}|^2\, |F_{q'q}|^2
\label{RFC.1c}
\end{equation}
and
\begin{equation}
R^{\mathrm{in},d}_{n'q' \to nq} = \Gamma_\mathrm{FC}\,
  f(E_{nq} - E_{n'q'} + eV_\mathrm{dot})
  \sum_\sigma |D^\sigma_{n'n}|^2\, |F_{q'q}|^2 .
\label{RFC.1d}
\end{equation}

For the tunneling barrier between the dot and ground we take the rates
\begin{equation}
R_\rightarrow = \frac{\gamma_\mathrm{barrier}\, eV_\mathrm{dot}}
  {\exp(eV_\mathrm{dot}/k_BT) - 1}
\end{equation}
for an electron tunneling from the dot to ground, and
\begin{equation}
R_\leftarrow = \frac{\gamma_\mathrm{barrier}\, eV_\mathrm{dot}}
  {1 - \exp(-eV_\mathrm{dot}/k_BT)}
\end{equation}
for the reverse process. Here,
$\gamma_\mathrm{barrier}$ is a bare tunneling rate. Note that this ansatz
satisfies detailed balance and leads to ohmic behavior since
the current is $I_\mathrm{barrier} = -e R_\rightarrow + e R_\leftarrow
  = e^2 \gamma_\mathrm{barrier} V_\mathrm{dot}$.

The Hamiltonians for the two auxiliary molecules describe a single orbital
with strong Coulomb repulsion but no coupling to a vibrational mode,
\begin{equation}
H_{\mathrm{aux}\:\nu} = \sum_\sigma (\epsilon_{\mathrm{aux}\:\nu} -
  eV_{\mathrm{aux}\:\nu})\,
  d^\dagger_\sigma d_\sigma
  + U_{\mathrm{aux}\:\nu}\, d^\dagger_\uparrow d_\uparrow d^\dagger_\downarrow
  d_\downarrow ,
\label{Haux.1}
\end{equation}
$\nu=1,2$, with eigenenergies $E^{\mathrm{aux}\:\nu}_n =
(\epsilon_{\mathrm{aux}\:\nu}
- eV_{\mathrm{aux}\:\nu})\, n_d + (U_{\mathrm{aux}\:\nu}/2) \, n_d(n_d-1)$.
The sequential-tunneling rates are of the same form as in (\ref{RFC.1a}),
(\ref{RFC.1b}), (\ref{RFC.1c}), and (\ref{RFC.1d}), with the FC matrix elements
replaced by unity and the relevant parameters shown in figures
\ref{fig.details}(b) and \ref{fig.details}(c).

Finally, the on-site potentials at the positions of the three molecules and of
the dot are obtained by solving the equations
\begin{eqnarray}
V_\mathrm{FC} & = & \frac{C_{\mathrm{FC},s} V_\mathrm{bias\:1}
  + C_\mathrm{\mathrm{FC},d} V_\mathrm{dot}
  + C_\mathrm{\mathrm{FC},g} V_\mathrm{in}}
  {C_{\mathrm{FC},s} + C_\mathrm{\mathrm{FC},d} + C_\mathrm{\mathrm{FC},g}} ,
\label{VC.1a} \\
V_{\mathrm{aux}\:1} & = & \frac{C_{\mathrm{aux}\:1,s} V_\mathrm{bias\:2}
  + C_{\mathrm{aux}\:1,d} V_\mathrm{out}
  + C_{\mathrm{aux}\:1,g} V_\mathrm{dot}}
  {C_{\mathrm{aux}\:1,s} + C_{\mathrm{aux}\:1,d}
  + C_\mathrm{\mathrm{aux}\:1,g}} , \\
V_{\mathrm{aux}\:2} & = & \frac{C_{\mathrm{aux}\:2,s} V_\mathrm{bias\:2}
  + C_{\mathrm{aux}\:2,d} V_\mathrm{in}
  + C_{\mathrm{aux}\:2,g} V_\mathrm{dot}}
  {C_{\mathrm{aux}\:2,s} + C_{\mathrm{aux}\:2,d}
  + C_\mathrm{\mathrm{aux}\:2,g}} , \\
V_\mathrm{dot} & = & \frac{Q_\mathrm{dot} + C_{\mathrm{FC},d} V_\mathrm{FC}
  + C_{\mathrm{aux}\:1,g} V_{\mathrm{aux}\:1}
  + C_{\mathrm{aux}\:2,g} V_{\mathrm{aux}\:2}}
  {C_{\mathrm{FC},d} + C_{\mathrm{aux}\:1,g} + C_{\mathrm{aux}\:2,g}
  + C_\mathrm{barrier}} ,
\label{VC.1d}
\end{eqnarray}
where $Q_\mathrm{dot} \equiv -e N_\mathrm{dot}$ is the charge on the dot. Note
that we do not include the charges on the molecules explicitly in these
equations because we have already done so in the Hamiltonians $H_\mathrm{FC}$,
$H_{\mathrm{aux}\:1}$, and $H_{\mathrm{aux}\:2}$.

\subsection{Monte Carlo simulations}

We are interested in the time-dependent behavior of the circuit. While the
master equation can be used to study dynamics
\cite{KoO05,KRO05,KOA06,TiE06,TiV12,DoT12,DYG12,SKB12}, it describes the
dynamics of \emph{ensembles}. We can gain more
insight by considering typical time evolutions of a \emph{single} system. For a
molecular transistor in the FC regime, this has been done in
\cite{KoO05,KRO05,KOA06}. We obtain the time evolution by performing Monte Carlo
simulations for the full circuit, using the transition rates given above.

We employ real-time Monte
Carlo simulations, which do not involve discretization of time. The state of
the circuit is characterized by the state $|nq\rangle$ of the FC molecule, the
states $|n\rangle_1$, $|n\rangle_2$ of the two auxiliary molecular transistors,
and the occupation number $N_\mathrm{dot}$ of the dot. The harmonic-oscillator
ladder is truncated at $q_\mathrm{max}=30$, which does not affect
the results since this value is never reached. At every step, we
first calculate the transition rates from this state to all other possible
states. The transition that actually happens is then selected pseudo-randomly
with a probability given by its branching fraction, i.e., its rate divided by
the sum of all rates (the total transition rate). The waiting time is drawn
pseudo-randomly from an exponential distribution with mean given by the inverse
of the total transition rate.

To avoid inessential complications, we assume a constant incoming current
$I_\mathrm{in}$, which should be valid as long as this current is large
compared to the additional current through the second auxiliary molecule. We
also assume that the output line is current free, $I_\mathrm{out}=0$. Despite
these simplifications, this is
still a complicated model due to the time dependence introduced by the
\textit{RC} elements. For the output line we have Kirchhoff's
current law
\begin{equation}
-e \dot N_\mathrm{out} = \frac{V_\mathrm{out}}{R_\mathrm{out}}
  + C_\mathrm{out} \dot V_\mathrm{out} ,
\label{Khout.1}
\end{equation}
where $N_\mathrm{out}(t)$ is the total number of electrons inserted from the
first auxiliary molecule into its drain electrode since an arbitrary reference
time. $N_\mathrm{out}(t)$ decreases if electrons flow into the device. If the
time of the tunneling process is small compared to the other time scales of the
circuit, the time derivative $\dot N_\mathrm{out}$ can be treated as a series
of $\delta$ functions. The solution of (\ref{Khout.1}) is
\begin{equation}
V_\mathrm{out}(t) = V_\mathrm{out}(0)\, e^{-t/\tau_\mathrm{out}}
  - \frac{e}{C_\mathrm{out}}\, e^{-t/\tau_\mathrm{out}}
  \int_0^t dt'\, e^{t'/\tau_\mathrm{out}}\, \dot N_\mathrm{out}(t') ,
\end{equation}
where $\tau_\mathrm{out} \equiv R_\mathrm{out} C_\mathrm{out}$ is the
characteristic time of the $RC$ element. For the input line, the current law
reads
\begin{equation}
I_\mathrm{in} - e \dot N_\mathrm{in} = \frac{V_\mathrm{in}}{R_\mathrm{in}}
  + C_{\mathrm{FC},g}\, (\dot V_\mathrm{in} - \dot V_\mathrm{FC}) .
\end{equation}
Here, the onsite potential of the FC molecule, $V_\mathrm{FC}$, depends on time
not only through the gate potential $V_\mathrm{in}$ but also through the
potential on the dot, $V_\mathrm{dot}$. The effect of the fluctuating dot
potential on $V_\mathrm{in}$ via the FC molecule is smaller than the
contribution from the current $- e \dot N_\mathrm{in}$ and we neglect it for
simplicity. We can then write
\begin{equation}
\dot V_\mathrm{FC} \approx \frac{\partial V_\mathrm{FC}}{\partial
  V_\mathrm{in}}\, \dot V_\mathrm{in} ,
\end{equation}
where the partial derivative can be obtained in terms of the capacitances
from the solution of (\ref{VC.1a})--(\ref{VC.1d}). We thus find
\begin{equation}
I_\mathrm{in} - e \dot N_\mathrm{in} \approx \frac{V_\mathrm{in}}{R_\mathrm{in}}
  + C_{\mathrm{FC},g}\, \left( 1 - \frac{\partial V_\mathrm{FC}}{\partial
  V_\mathrm{in}}\right)\, \dot V_\mathrm{in}
  \equiv \frac{V_\mathrm{in}}{R_\mathrm{in}}
  + C_\mathrm{in}\, \dot V_\mathrm{in} ,
\label{Iin.2}
\end{equation}
where we have defined an effective input capacitance $C_\mathrm{in}$. The
solution is then
\begin{equation}
\fl\qquad V_\mathrm{in}(t) = R_\mathrm{in} I_\mathrm{in}
  + [ V_\mathrm{in}(0) - R_\mathrm{in} I_\mathrm{in} ]\,
  e^{-t/\tau_\mathrm{in}}
  - \frac{e}{C_\mathrm{in}}\, e^{-t/\tau_\mathrm{in}}
  \int_0^t dt'\, e^{t'/\tau_\mathrm{in}}\, \dot N_\mathrm{in}(t') ,
\end{equation}
with $\tau_\mathrm{in} \equiv R_\mathrm{in} C_\mathrm{in}$.

The time dependences of $V_\mathrm{in}$ and $V_\mathrm{out}$ imply that the
transition rates change continuously in time and
not only discretely at the time of tunneling transitions. It would be difficult
to implement these time-dependent rates in the continuous-time approach.
However, this is not necessary since the characteristic times $\tau_\mathrm{in}$
and $\tau_\mathrm{out}$ are long compared to the waiting time between tunneling
events. Electrons are rapidly tunneling back and forth between the quantum dot
and ground, with a characteristic time of $1/\gamma_\mathrm{barrier}$.
Since $\tau_\mathrm{in},\tau_\mathrm{out}\gg 1/\gamma_\mathrm{barrier}$ for our
parameters, we can ignore the change of the rates during the Monte Carlo steps.

Consequently, for a Monte Carlo step of duration $\Delta t$, the output voltage
is updated according to
\begin{equation}
V_\mathrm{out}(t+\Delta t) = V_\mathrm{out}(t)\, e^{-\Delta t/\tau_\mathrm{out}}
  - \frac{e}{C_\mathrm{out}}\, \Delta N_\mathrm{out} ,
\end{equation}
where $\Delta N_\mathrm{out} \in \{-1,0,1\}$ is the number of electrons
inserted into the drain in this step. Similarly, the input voltage is updated
according to
\begin{equation}
V_\mathrm{in}(t+\Delta t) = R_\mathrm{in} I_\mathrm{in}
  + [ V_\mathrm{in}(t) - R_\mathrm{in} I_\mathrm{in} ]\,
  e^{-\Delta t/\tau_\mathrm{in}}
  - \frac{e}{C_\mathrm{in}}\, \Delta N_\mathrm{in} .
\end{equation}
All the other voltages are then recalculated using
(\ref{VC.1a})--(\ref{VC.1d}) and the algorithm loops back to the calculations of
the transition rates out of the new state. The Monte Carlo step is repeated
until the simulated time $t$ exceeds an equilibration time $t_\mathrm{init}$ in
order to get rid of any dependence on the initial state.
Then, the simulated time is reset to zero and the state of the system and the
various voltages are recorded at each Monte Carlo step until the simulated time
exceeds $t_\mathrm{sim}$.

\section{Results and discussion}
\label{sec.results}

A functional device is obtained for the following parameters: The energies
in the molecular Hamiltonians are expressed in terms of a basic energy unit
$\epsilon_0$ as
\begin{eqnarray}
\epsilon_\mathrm{FC} = 12.9\, \epsilon_0 , \quad
U_\mathrm{FC} = 500\, \epsilon_0 , \quad
\hbar\omega_v = \epsilon_0 , \quad
\lambda = 4 , \\
\epsilon_{\mathrm{aux}\:1} = -3.6\, \epsilon_0 , \quad
\epsilon_{\mathrm{aux}\:2} = -5.5\, \epsilon_0 , \quad
U_{\mathrm{aux}\:1} = U_{\mathrm{aux}\:2} = 500\, \epsilon_0 ,
\end{eqnarray}
where the values for the Coulomb repulsions prevent double occupancy and are
thus effectively infinite. The temperature is taken to be $k_BT =
0.05\,\epsilon_0$. The parameters for the FC molecule are thus essentially the
same as used for the Monte Carlo simulations in \cite{KoO05,KRO05}. In
particular, the electron-vibron coupling $\lambda=4$ is strong and the molecule
should show FC blockade. The other parameters of the circuit are
\begin{eqnarray}
V_\mathrm{bias\:1} & = & - V_\mathrm{bias\:2} = -3\, \epsilon_0/e , \\
C_{\mathrm{FC},s} & = & C_{\mathrm{FC},d} = C_{\mathrm{FC},g} = C_0 , \\
C_{\mathrm{aux}\:1,s} & = & C_{\mathrm{aux}\:1,d}
  = C_{\mathrm{aux}\:2,s} = C_{\mathrm{aux}\:2,d} = 0.03\, C_0 , \\
C_{\mathrm{aux}\:1,g} & = & C_{\mathrm{aux}\:2,g} = 0.1\, C_0 , \\
C_\mathrm{barrier} & = & C_0 , \\
C_\mathrm{out} & = & 0.6\, C_0 , \\
R_\mathrm{in} & = & R_\mathrm{out} = 16667\, R_0 , \\
\Gamma_\mathrm{FC} & = & \Gamma_0 , \\
\Gamma_{\mathrm{aux}\:1} & = & \Gamma_{\mathrm{aux}\:2} = 2\, \Gamma_0 , \\
\gamma_\mathrm{barrier} & = & 0.01\, \Gamma_0 .
\end{eqnarray}
Note that the contacts of each molecule to its source and drain electrodes are
symmetric. This symmetry is not essential for the operation of the
electronic neuron. For the given capacitances, the
effective input capacitance, defined in (\ref{Iin.2}), is $C_\mathrm{in} =
0.599\, C_0$, which is why we have chosen $C_\mathrm{out} = 0.6\, C_0$.

We measure time in units of the inverse rate, $\Gamma_0^{-1}$. To have a
consistent unit system, we require $R_0 C_0 = \Gamma_0^{-1}$. If we choose a
time scale of $\Gamma_0^{-1} = 10^{-10}\,\mathrm{s}$, which leads to feasible
tunneling times \cite{PPL00,HGF06}, and a unit capacitance of
$10^{-19}\,\mathrm{F}$, which appears to be realistic
\cite{RZM97,NaW03,BeS09,WuY10}, we obtain $R_0 = \Gamma_0^{-1}/C_0 =
1\,\mathrm{G}\Omega$. This resistance is much larger than the
quantum resistance $R_K = h/e^2 = 2.58\times 10^4\,\Omega$ of an open channel.
Moreover, the input and output resistances in our circuit are much larger than
$R_0$, which may be problematic. However, larger capacitances or faster
tunneling would allow us to use smaller resistances. Our unit of energy
is the energy quantum of the vibrational
mode, $\epsilon_0 = \hbar\omega_v$. $\epsilon_0 = 30\,\mathrm{meV}$ is a
reasonable order of magnitude \cite{PPL00}. Since we will see that typical
voltage signals are on the order of a few times $\epsilon_0/e$, this leads to
voltages on the order of $0.1\,\mathrm{V}$.
As noted above, in the simulations we assume a constant input current
$I_\mathrm{in}$, given in units of $e\Gamma_0^{-1}$, and a vanishing output
current $I_\mathrm{out}=0$. The largest input current we use is $|I_\mathrm{in}|
= 6\times 10^{-4}\, e\Gamma_0^{-1}$, which small in natural units.
We first equilibrate the system for a time
$t_\mathrm{init} = 5\times 10^6\,\Gamma_0^{-1}$, and then record observables for
$t_\mathrm{sim} = 10^7\, \Gamma_0^{-1}$, unless noted otherwise.

\begin{figure}
\centerline{\includegraphics[scale=0.4,clip]{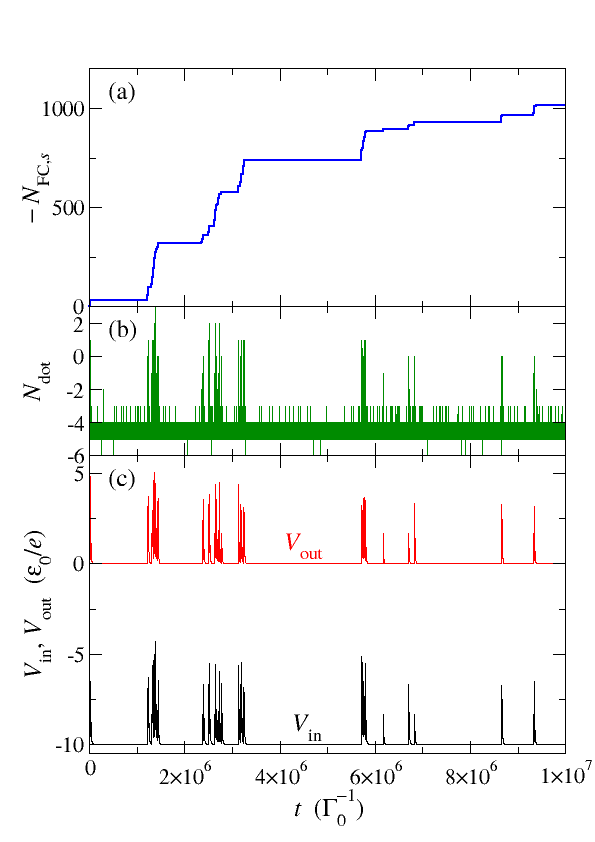}}
\caption{Simulation of the molecular neuron in the ON state for the
input current $I_\mathrm{in} = -6\times 10^{-4}\,e\Gamma_0^{-1}$.
The other parameters are given in the text. (a)
Change in the number of electrons in the source electrode of the FC molecule,
$N_{\mathrm{FC},s}$. (b) Number $N_\mathrm{dot}$ of electrons in the quantum
dot. (c) Voltages $V_\mathrm{in}$ and $V_\mathrm{out}$ measured at the input and
output contacts of the neuron, respectively.\label{fig.on}}
\end{figure}

Results for the ON state are given in figure \ref{fig.on}. Here, the input
current is $I_\mathrm{in} = -6\times 10^{-4}\,e\Gamma_0^{-1}$. Figure
\ref{fig.on}(a) shows the change in the number of electrons in the source
electrode of the FC molecule, $N_{\mathrm{FC},s}$. Clearly, the electrons flow
in avalanches, typical for the FC regime. Comparison with figure 3 of
\cite{KoO05} shows that the auxiliary molecular transistors do not appreciably
change this behavior. In particular, feedback from the current through the
second auxiliary molecule does not choke off the avalanche-like transport.

In figure \ref{fig.on}(b) we plot the number $N_\mathrm{dot}$ of electrons in
the quantum dot, relative to the neutral state. Negative electron numbers simply
mean that the dot is positively charged. The rapid fluctuations are due to
electrons tunneling through the barrier between the dot and ground. The
electron number preferentially assumes the values $-4$ and $-5$ since these
values tune the onsite potential $V_\mathrm{dot}$ close to zero, which is the
value it would assume in equilibrium with only the ground contact.
The avalanches transmitted through the FC
molecule are clearly visible. Note that the dot charge stays relatively small
so that a large organic molecule or small metallic cluster should be able to
accomodate it.

Figure \ref{fig.on}(c) shows the input and output voltages, $V_\mathrm{in}$ and
$V_\mathrm{out}$, respectively. $V_\mathrm{in}$ contains a constant term
resulting from the constant input current. Sharp spikes are evident in both
voltages, showing that the active neuron indeed fires voltage spikes into
\emph{both} the input \emph{and} the output line.

\begin{figure}
\centerline{\includegraphics[scale=0.4,clip]{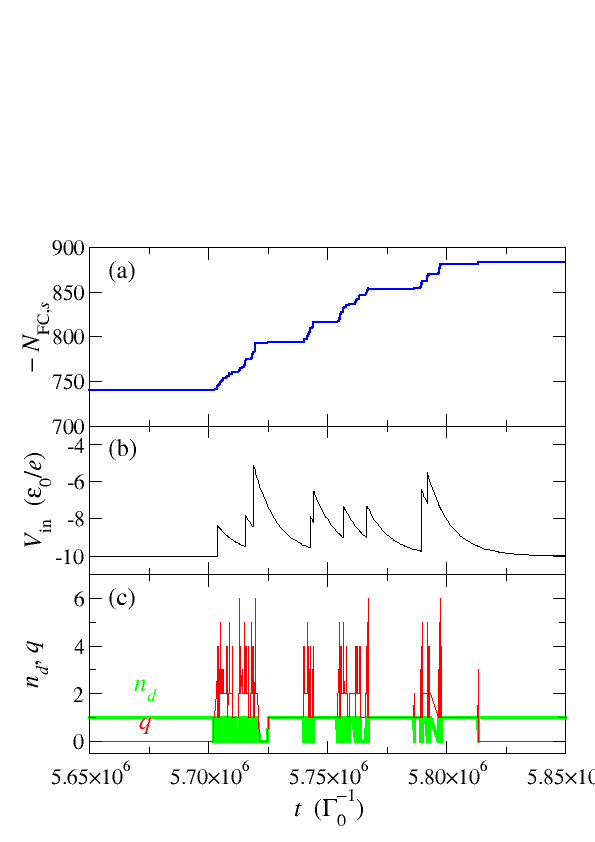}}
\caption{Details of the simulation presented in figure \ref{fig.on} for a
shorter time interval. (a) Change in the number of electrons in the source
electrode of the FC molecule, $N_{\mathrm{FC},s}$. (b) Voltage $V_\mathrm{in}$
measured at the input contact of the neuron. This voltage acts as the gate
voltage of the FC molecular transistor. (c) Occupation number $n_d$ and
harmonic-oscillator quantum number $q$ of the FC molecule.\label{fig.on2}}
\end{figure}

The spiky voltage $V_\mathrm{in}$ is the gate voltage seen by the FC
molecular transistor. As we will see below, $V_\mathrm{in}$ reaches values that
would, if applied continuously, switch off the FC molecule.
Nevertheless, the transport through the FC molecule is
hardly affected by these spikes, as figure \ref{fig.on} shows. To
understand this, we focus on a single series of spikes, taken from the same data
as in figure \ref{fig.on}. Figures \ref{fig.on2}(a) and \ref{fig.on2}(b) show
$N_{\mathrm{FC},s}$ and $V_\mathrm{in}$, respectively, for a shorter time
interval. In figure \ref{fig.on2}(c) we plot the occupation number $n_d$ and
the harmonic-oscillator quantum number $q$ of the FC molecule. Evidently,
$q\ge 1$ during most of the episode. Thus the FC molecular
transistor stores energy in the vibrational mode during the avalanches, which
allows it to stay active even though the gate voltage is strongly reduced in
magnitude. We see that the \emph{memristive} properties of the FC molecule are
important: The current flowing through it does not only depend on the
instantaneous gate and bias voltages but also on its history, here realized by
the vibrational quantum number $q$~\cite{review}.

\begin{figure}
\centerline{\includegraphics[scale=0.4,clip]{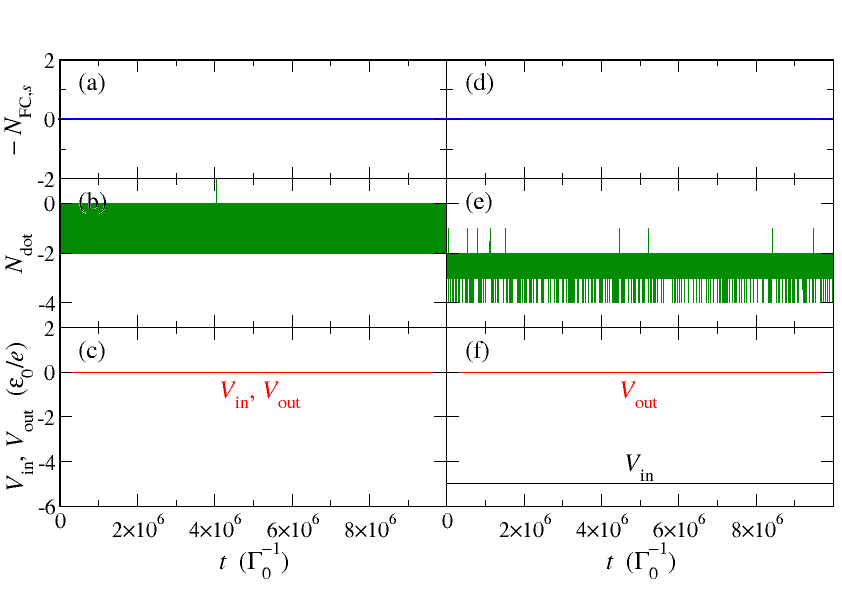}}
\caption{(a)--(c) Simulation of the molecular neuron in the OFF state
for vanishing input current, $I_\mathrm{in} = 0$. The other parameters are the
same as for figure \ref{fig.on}. (a) Change in
the number of electrons in the source electrode of the FC molecule,
$N_{\mathrm{FC},s}$. (b) Number $N_\mathrm{dot}$ of electrons in the quantum
dot. (c) Voltages $V_\mathrm{in}$ and $V_\mathrm{out}$ measured at the input and
output contacts of the neuron, respectively.
(d)--(f) The same quantities obtained for the input current
$I_\mathrm{in} = -3\times 10^{-4}\,e\Gamma_0^{-1}$. The other parameters are the
same as for figure \ref{fig.on}. The neuron is still inactive.\label{fig.off}}
\end{figure}

Results for the OFF state are shown in figures \ref{fig.off}(a)--(c) for
$I_\mathrm{in} = 0$. The same quantities as for the ON state in figure
\ref{fig.on} are plotted. The FC molecule is not transmitting. The quantum dot
only shows the fluctuating occupation number due to tunneling through the
barrier, which is not sufficient to switch on the auxiliary molecular
transistors.
We also require the molecular neuron to show \emph{threshold behavior}---it
should not fire spikes for input currents below a certain threshold. Figures
\ref{fig.off}(d)--(f) show that the device can indeed take a finite input
current without becoming active. Here, we have used $I_\mathrm{in} =
-3\,e\Gamma_0^{-1}$, half the value for the active state in figure~\ref{fig.on}.

\begin{figure}
\centerline{\includegraphics[scale=0.35,clip]{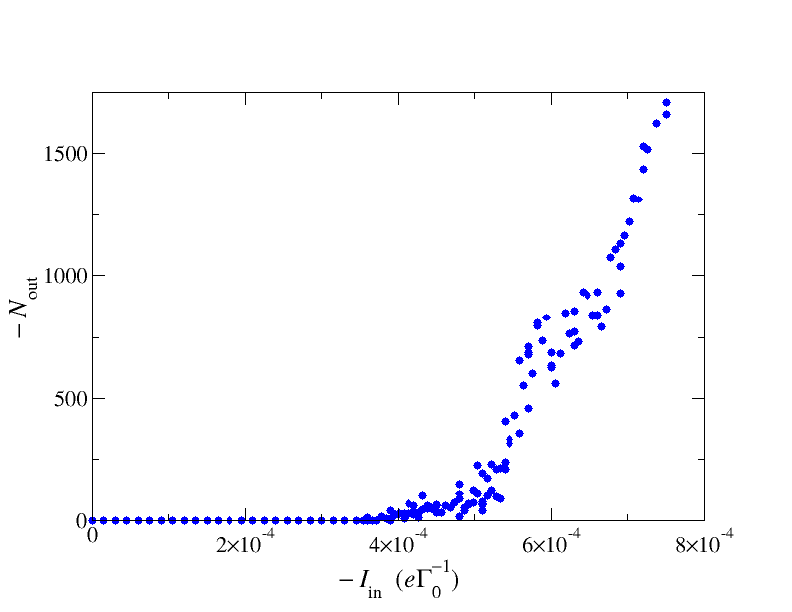}}
\caption{Number $N_\mathrm{out}$ of electrons inserted into the drain electrode
of the second auxiliary molecule during a simulation time of
$t_\mathrm{sim} = 5\times 10^7\, \Gamma_0^{-1}$, as a function of the input
current $I_\mathrm{in}$.\label{fig.switch}}
\end{figure}

For neural functionality, it is desirable that the switching between OFF
and ON states takes place over a narrow range of input currents. To study
this onset, we plot in figure \ref{fig.switch} the total number $N_\mathrm{out}$
of electrons inserted into
the drain electrode of the second auxiliary molecule, as a function of the
input current $I_\mathrm{in}$. A longer
simulation time of $t_\mathrm{sim} = 5\times 10^7\, \Gamma_0^{-1}$ has been
used. There is considerable noise, in particular for intermediate currents,
since the FC molecule transmits electrons in avalanches. Between avalanches,
the FC molecule returns to the vibrational ground state, $q=0$, and thus does
not retain any memory of the previous avalanche. Figure \ref{fig.switch}
essentially shows the shot noise of independent avalanches. Analysis
of the time series (not shown) indicates that the main effect of tuning
$I_\mathrm{in}$ is to change the delay times between avalanches, not so much
their duration. In any case, it is clear that the neuron is inactive over a
wide range of input currents.

\section{Summary and conclusions}
\label{sec.summary}

We have proposed a design for a molecular-electronics realization of a neuron.
The critical component of this artificial neuron is a molecular transistor with
strong electron-vibration
coupling, which is tuned to the FC-blockade regime when the neuron is active.
In this regime, electrons flow in avalanches. The charge dumped by an
avalanche into a small quantum dot or large molecule is used to activate two
auxiliary molecular transistors, which lead to voltage pulses traveling along
the output line and, importantly, also back up the input line. Employing
Monte Carlo simulations for the dynamics of the circuit within the
sequential-tunneling approximation, we have demonstrated that the device shows
the desired neural behavior. Such a system can be fabricated with experimental
capabilities available now or in the near future. It can be used for purposes
other than the one discussed here. For instance,
with an appropriate choice of system parameters it can operate as an amplifier.
Most importantly, it can be used for dynamic storage and transfer of information
in complex and highly integrated artificial networks that mimic the behavior of
biological neural systems.

\ack


Financial support by the Deutsche Forschungsgemeinschaft, through Research Unit
1154, \textit{Towards Molecular Spintronics}, and by the US National Science
Foundation grant No.\ DMR-0802830 is gratefully acknowledged.

\end{document}